\documentclass[conference]{IEEEtran}
\IEEEoverridecommandlockouts
\usepackage{cite}
\usepackage{amsmath,amssymb,amsfonts}
\usepackage{algorithmic}
\usepackage{graphicx}
\usepackage{epsfig}
\usepackage{textcomp}
\usepackage{graphicx}
\usepackage{subcaption}

\def\BibTeX{{\rm B\kern-.05em{\sc i\kern-.025em b}\kern-.08em
    T\kern-.1667em\lower.7ex\hbox{E}\kern-.125emX}}
\begin{document}

\title{AFP-Net: Realtime Anchor-Free Polyp Detection in Colonoscopy \\
{\footnotesize }
\thanks{``*" indicates equal contribution.}
}

\author{\IEEEauthorblockN{ Dechun Wang*, Ning Zhang*, Xinzi Sun, Pengfei Zhang, Chenxi Zhang, Yu Cao, Benyuan Liu}
\IEEEauthorblockA{\textit{Department Of Computer Science} \\
\textit{University of Massachusetts Lowell}\\
Lowell, MA \\
\{dechun\_wang, ning\_zhang, xinzi\_sun, pengfei\_zhang,  chenxi\_zhang\}@student.uml.edu\\ \{ycao, bliu\}@cs.uml.edu}}
\IEEEpubid{\makebox[\columnwidth]{\copyright~2019 IEEE \hfill} \hspace{\columnsep}\makebox[\columnwidth]{ }}

\maketitle
\begin{abstract}
Colorectal cancer (CRC) is a common and lethal disease. Globally, CRC is the third most commonly diagnosed cancer in males and the second in females. For colorectal cancer, the best screening test available is the colonoscopy. During a colonoscopic procedure, a tiny camera at the tip of the endoscope generates a video of the internal mucosa of the colon. The video data are displayed on a monitor for the physician to examine the lining of the entire colon and check for colorectal polyps. Detection and removal of colorectal polyps are associated with a reduction in mortality from colorectal cancer. 
However, the miss rate of polyp detection during colonoscopy procedure is often high even for very experienced physicians. The reason lies in the high variation of polyp in terms of shape, size, textural, color and illumination. Though challenging, with the great advances in object detection techniques, automated polyp detection still demonstrates a great potential in reducing the false negative rate while maintaining a high precision. In this paper, we propose a novel anchor free polyp detector that can localize polyps without using predefined anchor boxes. To further strengthen the model, we leverage a Context Enhancement Module and Cosine Ground truth Projection. Our approach can respond in real time while achieving state-of-the-art performance with 99.36\% precision and 96.44\% recall.		
\end{abstract}

\section{Introduction}
Colorectal Cancer (CRC) is the third most common cancer worldwide and the second most lethal cancer in USA\cite{doi:10.3322/caac.21395}, with an estimate of 1.8 million new diagnosed colorectal cancer cases and 881,000 deaths in 2018\cite{doi:10.3322/caac.21492}. Colorectal cancer often begins with the growth of tissues known as polyps. Most of these polyps are initially benign but some of them will become malignant over time. Early diagnosis via colonoscopy is widely believed to be the best way for the prevention of colorectal cancer. During a colonoscopy, specialized physicians will carefully inspect the intestinal wall. However, the operations can have a false negative rate of 25\% \cite{leufkens_oijen_vleggaar_siersema_2012}. Multiple independent factors contribute to the miss rate. Some of the factors are related to the polyp appearance, eg. a small and flat polyp is less likely to be perceived, while some other causes can be operational: the camera is moving too fast, leading physicians to fail to catch the suspicious regions which require more efforts. For these reasons, real time automated polyp detection can act as a complementary tool to assist physicians in improving the sensitivity of the diagnosis. 
\begin{figure*}
	\begin{center}
		\includegraphics[width=0.8\linewidth,height=2in]{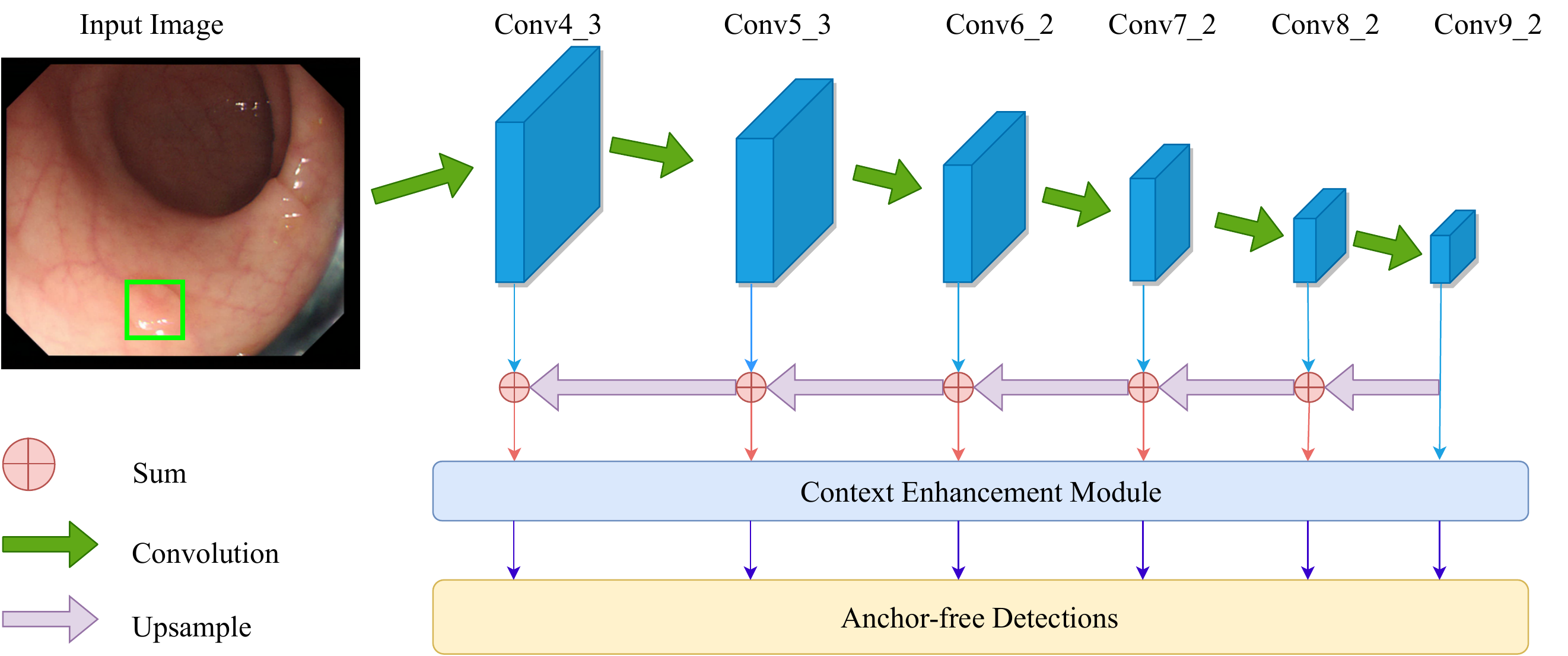}
	\end{center}
	\caption{The network architecture of our proposed framework. The network includes a top-down up-sampling path to enrich semantic features at different scales, followed by a Context Enhancement Module (CEM) to increase the receptive field. The enhanced feature maps are then fed into Anchor-free detection heads to predict bounding boxes.} 
	\label{fig:network}
\end{figure*}

In the past, automated polyp detection mainly utilized hand-crafted color, shape and texture features to distinguish polyps from normal mucosae \cite{inproceedings,4379193}. Bernal et. al \cite{cvc_clinic} used the appearance of the polyps as a key factor. They assumed all polyps have protruding surface and used valley detection as the feature extractor. However, polyp and normal mucosae can share similar features in edge and texture when polyps have flat surfaces as normal mucosae. In addition, some mucosae with vally-reach structures can appear similar to polyps. All these facts result in poor performance for these approaches in real applications.  

Recently, Convolutional Neural Network (CNN) based object detectors such as SSD \cite{DBLP:SSD} and Faster R-CNN \cite{DBLP:Faster-RCNN} dominate the object detection task across different image modalities. Compared with traditional approaches where features are fully designed based on human knowledge, these CNN based detectors can learn rich features automatically in the deep backbone networks. This is the key contributing factor for deep learning to thrive \cite{Zhang:2017:IMR:3123266.3123332,7545830,ALCANTARA201766,7545842,7255222,sun2019People}. 

To make it more clear, we divide CNN based detectors into the backbone part (eg. ResNet \cite{he2016deep}, VGG \cite{vgg16}) and the head part which is more specific to individual detectors. For the detection head part, the ``anchor" concept is shared by both single-stage \cite{DBLP:SSD, DBLP:YOLO, lin2017feature} and two-stage \cite{he2017mask, DBLP:Faster-RCNN} detectors. Anchors represent different box templates (at different scales). This strategy works by enumerating box templates of different scales and aspect ratios with respect to the dataset in process. In other words, anchors should adapt to the characteristics of the dataset. However, designing anchor sets and assigning objects to specific anchors require extensive experience. It is also indicated in \cite{eggert2017closer} that the choice of anchors is especially important for small objects. Moreover, when we assign objects to certain anchors, IoU is usually the major criterion. Different IoU thresholds can result in significant performance variations. 



Motivated by these observations, the ``Anchor-Free" approaches \cite{DBLP:cornernet,DBLP:journals/corr/abs-1904-08189, wang2019region, zhou2019keypoint,zhang20193d} have received much attention recently, where the anchor mechanism is removed and objects are represented as keypoints. For instance, CornerNet \cite{DBLP:cornernet} represents an object as a pair of keypoints (top-left and bottom-right corners). To better detect these corners, a special corner pooling is devised to enrich the context information. Different from CornerNet, Zhou et. al \cite{zhou2019keypoint} represent an object as a single keypoint located in the center. In this way, the time-consuming pairing is removed and the model achieves the best trade-off between performance and inference speed. 

In the polyp detection, we observe that the objects (polyps) under concern actually do not overlap much with each other and the shapes do not vary much either. Therefore, we believe the idea of ``objects as keypoints" fits well with our application scenario. Moreover, removing the anchor mechanism can reduce the parameters in detection heads thus result in fewer highly overlapped proposals during inference, which can potentially accelerate the inference speed. 


In this paper, we propose a novel anchor-free detector for fast polyp detection. 
Similar to \cite{zhou2019keypoint}, we formulate objects as center points, yet removing the time-consuming center pooling in CenterNet \cite{DBLP:journals/corr/abs-1904-08189} to achieve real time response. The role of the center pooling is replaced by a context enhancement module as well as a feature pyramid design. In addition, we devise a special cosine ground-truth projection strategy to compensate for the potential drop in recall caused by the removal of the anchor mechanism.  
Our proposed polyp detector outperforms previous studies and achieves the state of the art performance in terms of both accuracy and inference speed. 


\section{Related Work}
\subsection{Polyp Detection}
With the success of deep learning in natural image processing, CNN based polyp detectors were proposed in the last few years. Compared to hand-crafted features, CNN based detectors automate the process of extracting abstract and discriminative features. They are more robust and require less domain knowledge, making them particularly suitable for this task. As our approach is also CNN based, we only discuss with details CNN based detectors in this paper.

Zhang et. al \cite{rest-yolo} proposed a two-step pipeline for polyp detection in endoscopic videos. In the first step, they use a pre-trained ResYOLO to detect suspicious polyps. The polyps are assumed to be stable without a sudden move from one location to another between two consecutive frames. Therefore, in the second step, a Discriminative Correlation Filter based tracking approach was proposed to leverage the temporal information. This tracking based method refines the detection results given by ResYOLO and is capable of locating polyps missed by the detector in consecutive frames.  

Mohammed et. al \cite{DBLP:y-net} proposed Y-Net for this task. It consists of two fully convolutional encoders followed by a fully convolutional decoder. The motivation is that a model pre-trained on natural images may not generalize well on medical images. To mitigate performance degradation due to the domain-shift (natural images to medical images), they slowly fine-tune the first encoder from the pre-trained network while aggressively train from scratch for the second encoder. Two encoders are of the same network architecture and the outputs are combined with a sum-skip-concatenation connection before coming to the decoder network.

However, seemly contradicting with \cite{DBLP:y-net}, Mo et. al \cite{polyp-faster-rcnn} proved that fine-tuning a pre-trained Faster R-CNN can work considerably well. In addition, Shin et. al \cite{fatser-rcnn-with-post-learning} proposed a post learning scheme to enhance the Faster R-CNN detector. This post learning scheme automatically collects hard negative samples and retrains the network with selected polyp-like false positives, which functions similarly to boosting.

Our proposed method is the first one to apply the anchor-free approach to automated polyp detection. We believe that anchor-free design is a very practical solution in our case in terms of speed and accuracy. Unlike natural images where pre-defined anchors are introduced to attack the occlusion issue, in medical images such as Computed Tomography and Colonoscopy Images, occlusion is rare. Another concern for this polyp detection task is the real time requirement. With anchors, a large number of overlapped proposals would be proposed, putting significant pressure on the post-processing step (Non-Maximum Suppression). Therefore, the anchor free mechanism would fit better in this polyp detection task. 

\subsection{Anchor Free Detectors}
While almost all the state of the art object detectors employ pre-defined anchors, anchor-free object detectors \cite{DBLP:cornernet, DBLP:journals/corr/abs-1904-08189, wang2019region,tian2019fcos, kong2019foveabox} have received much attention in recent years because of their better adaptability towards different datasets. Representative approaches include CornerNet \cite{DBLP:cornernet} and CenterNet \cite{DBLP:journals/corr/abs-1904-08189}.

In CornerNet \cite{DBLP:cornernet}, objects are represented as pairs of keypoints: top-left corners and bottom-right corners. The network is trained to predict a heatmap for all top-left corners and for all bottom-right corners respectively in parallel. To associate them in pairs, for each corner an embedding vector is also learned: pushing away corners belonging to different objects while pulling together corners in the same group. To further strengthen the ``corner learning" process, the authors also devise a special Corner Pooling. This pooling consists of two separate one-directional pooling (horizontal and vertical) before combining them together at the end. In this way, more context information is gathered to identify corners. However, because of the pairing process, the inference speed drops significantly.

Zhou et. al \cite{zhou2019keypoint} use the center keypoint to represent an object. In this way, the burden of learning to group corners is removed and the model achieves the best performance-speed trade-off. Our work follows this idea. But we add more constraints when assigning the label to keypoints in the context of the feature pyramid. These constraints are later proved to be critical in our experiments. Moreover, as indicated in CornerNet, enriching context information plays a key role in the success of anchor-free detectors. We also explore in this direction by adding a Context Enhancement Module.


\section{Methodology}
In this section, we will describe our proposed anchor-free approach in detail including the network design and the associated technical components.

\subsection{Network Architecture}
Figure \ref{fig:network} illustrates our framework design. It is a fully convolutional network that classifies and localizes objects on each enhanced feature map. Our network uses VGG16 \cite{vgg16} as the backbone 
Our framework selects $k=6$ feature maps from the backbone: \textit{conv4\_3, conv5\_3, conv6\_2, conv7\_2, conv8\_2 } and \textit{conv9\_2}, responsible for detecting objects at different scales. In order to increase semantics information for each feature map, we build a feature pyramid similarly to FPN \cite{DBLP:FPN}. Note that we use Deconvolution (Transposed Convolution) for up-sampling instead of interpolation. Upstream and downstream features of the same scale are combined by element-wised addition. For each feature map $S_{i}$ where $i \in [0,k-1)$, we first use 1x1 convolution to smooth the upper feature map $S_{i+1}$, which is then up-sampled to the size of $S_{i}$. Finally, we add up-sampled $S_{i+1}$ to $S_{i}$. The enhanced feature map $E_{i}$ can be mathematically represented as follows: 
\begin{equation}
E_i = S_i + upsample(smooth(S_{i+1}))
\end{equation}

To further increase context information for small objects, we feed each feature map $E_i$ to a context module before forwarding them to the anchor-free detection heads. These detection heads are single-staged and have similar structures to the heads in SSD where two parallel subnets are dedicated for classification and localization respectively.  

\begin{figure}[t]
	\begin{center}
		\includegraphics[width=0.9\linewidth]{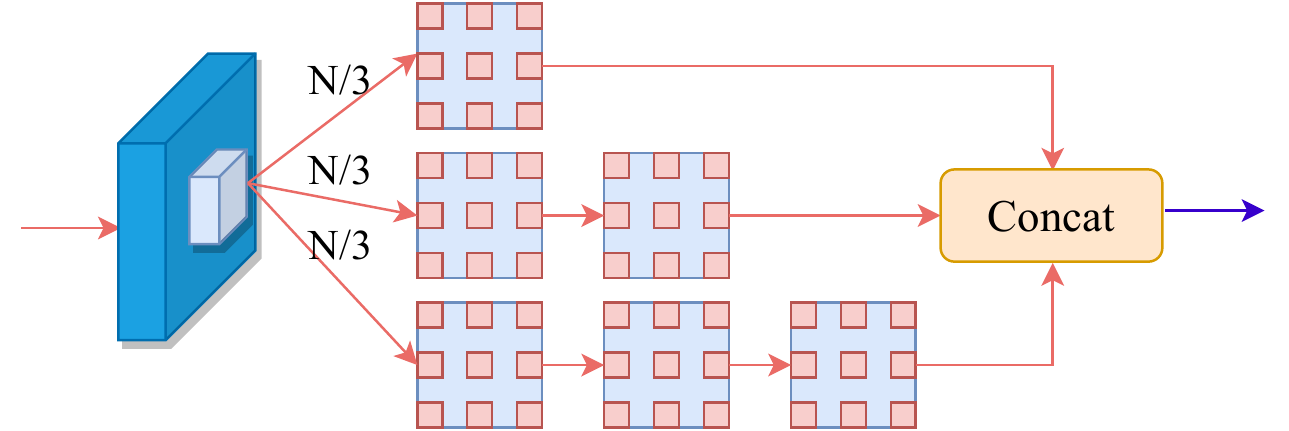}
	\end{center}
	\caption{Illustration of the context enhancement module. The input feature channels were first divided into three chunks, each fed into dilated convolution layers of different depths before they are concatenated. In this way, the effective receptive field is enlarged while detailed information is still retained.}
	\label{fig:context_module}
\end{figure}

\subsection{Context Enhancement Module (CEM)}
In order to increase context information for small objects, we apply a similar context enhancement module (illustrated in Fig. \ref{fig:context_module}) in \cite{DBLP:SSH,DBLP:DSFD}. Since our anchor-free detection heads are of a fully convolutional manner, increasing context information is equivalent to enlarge the receptive field of our detection heads. However, instead of using 5x5 or 7x7 filters to enlarge the receptive field, we adopt the dilated convolution following \cite{DBLP:DSFD,5x5_3x3}. One good property about dilated convolutions is that it brings in fewer parameters to achieve the same size of the receptive field. 
Note that the input channels are equally split into three branches with each branch being of different depths. Outputs of all branches are concatenated at the end. By employing this context enhancement module, we find in experiments that it can considerably increase the precision performance.
\begin{figure}[t]
	\centering
	\begin{subfigure}[b]{0.4\linewidth} %
		\centering
		\includegraphics[width=1\textwidth]{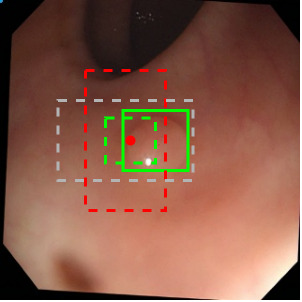}%
		\caption{Standard anchor based ground truth matching.}%
		\label{fig:anchor_compare-1}%
	\end{subfigure}%
	\quad%
	\begin{subfigure}[b]{0.4\linewidth}%
		\centering
		\includegraphics[width=1\textwidth]{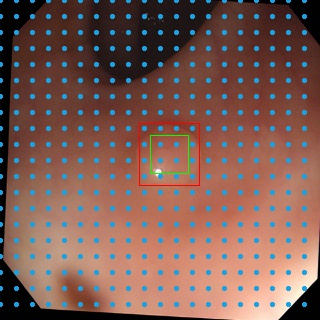}%
		\caption{Our anchor free ground truth matching. }%
		\label{fig:anchor_compare-2}%
	\end{subfigure}%
	\caption{Comparison between anchor-based detectors (a) and our anchor free detector (b) in ground truth target assigning.}
	\label{fig:anchor_compare} 
\end{figure}

\begin{figure*}[t]
	\centering
	\begin{subfigure}[b]{0.15\linewidth} 
		\centering
		\includegraphics[width=1\textwidth]{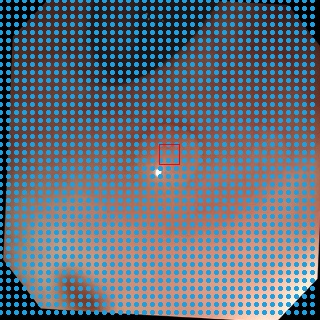}
		\caption{conv4\_3}
		\label{fig:projection-1}
	\end{subfigure}
	\begin{subfigure}[b]{0.15\linewidth}
		\centering
		\includegraphics[width=1\textwidth]{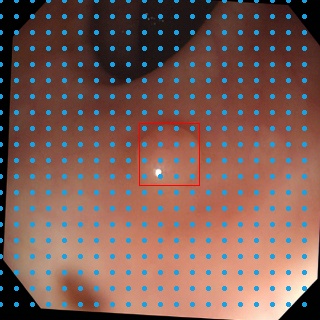}
		\caption{conv5\_3 }
		\label{fig:projection-2}
	\end{subfigure}
	\begin{subfigure}[b]{0.15\linewidth}
		\centering
		\includegraphics[width=1\textwidth]{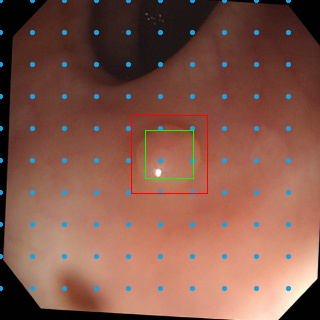}
		\caption{conv6\_2 }
		\label{fig:projection-3}
	\end{subfigure}
	\begin{subfigure}[b]{0.15\linewidth}
		\centering
		\includegraphics[width=1\textwidth]{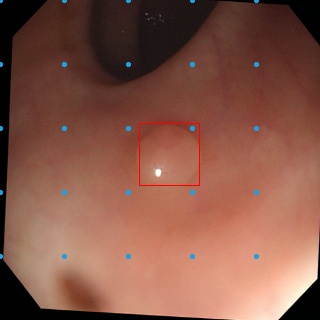}
		\caption{conv7\_2 }
		\label{fig:projection-4}
	\end{subfigure}
	\begin{subfigure}[b]{0.15\linewidth}
		\centering
		\includegraphics[width=1\textwidth]{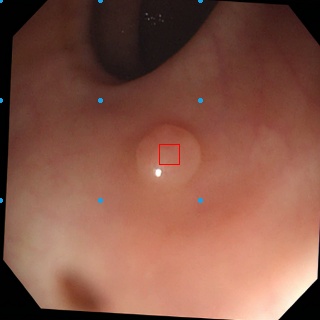}
		\caption{conv8\_2 }
		\label{fig:projection-5}
	\end{subfigure}
	\begin{subfigure}[b]{0.15\linewidth}
		\centering
		\includegraphics[width=1\textwidth]{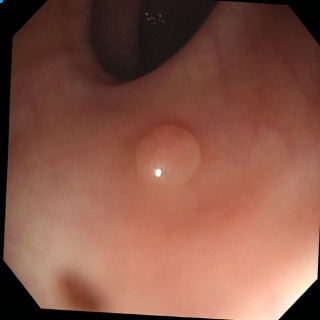}
		\caption{conv9\_2 }
		\label{fig:projection-6}
	\end{subfigure}
	\caption{Ground truth projection on different feature maps (a - f represent different scales). The blue dots are the projected points from the feature map. The green rectangles and red rectangles represent the positive and non-negative region respectively. In this figure, the ground-truth's was originally assigned to (c) conv6\_2. We project its non-negative region to different levels of feature map with reduced sizes. The amount of reduction is based on the distance to (c) conv6\_2.}
	\label{fig:projection} 
\end{figure*}

\subsection{Anchor-free Label Assignment}
We adopt a very different label assigning procedure in our anchor-free design compared with the anchor-based framework (Fig. \ref{fig:anchor_compare}). In anchor-based detectors (Fig. \ref{fig:anchor_compare-1}), the green dashed rectangle is a positive anchor with $IoU > 0.5$ and red dashed rectangle represents a negative anchor with $IoU < 0.3$. The white dashed anchor would be ignored as its $IoU$ is in between. Unlike anchor-based approaches relying on manually set up overlap thresholds, in our anchor-free design, center points are labeled solely based on the location and the size of ground truth boxes. 
The whole process is demonstrated in Fig. \ref{fig:anchor_compare-2}. The blue dotted grid represents center points. Any center points falling outside of the red rectangle, inside of the green rectangle and in the gap between the two will be assigned as negative, positive and ignored respectively. 

More formally, consider an object $b = (c_x , c_y, w, h)$, where $(c_x, c_y)$ is the center position of the bounding box; $ w,h $ are the width and height respectively. Suppose this object was assigned to an arbitrary feature map $K_l$ with size $m$ and stride $s_l$ during training, we have $m^2$ (we assume the image is a square for simplicity) center points $p_{ij}$ from set $\widehat{P} = \{x_i,y_j\}^{m}_{i,j=1}$ where $x_i = s_l \times i$ and $y_j = s_l \times j$. We define an positive region $b^l_p = (c_x,c_y,w^l_p,h^l_p)$ and non-negative region $b^l_n = (c_x,c_y,w^l_n,h^l_n)$. We mark center points as positive if $p_{ij} \in b^l_p $, negative if $p_{ij} \notin b^l_n$ and ignored if $p_{ij} \notin b^l_p \land p_{ij} \in b^l_n $. The size of the positive region and non-negative region is controlled by scale factors $\varepsilon_p$  and $\varepsilon_n$ in proportion to the ground truth, i.e. $w^l_p =  \varepsilon_p w ,h^l_p= \varepsilon_p h ,w^l_n =  \varepsilon_n w, h^l_n=\varepsilon_n h$. We use $ \varepsilon_p=0.75$ and $\varepsilon_n=1.25  $ in this paper.

\subsection{Cosine Ground-truth Projection}
It has been shown that distributing objects of different sizes to different scales can greatly improve the detector performance. In this way, objects will be detected at the best granularity with respect to their size. On the other hand, numerically, bounding box regression would also benefit from this technique as the loss could be better constrained. Almost all current state-of-the-art object detectors \cite{DBLP:SSD,DBLP:Faster-RCNN,DBLP:YOLO,DBLP:cornernet,DBLP:FPN} employ this strategy. We also adopt this multi-scale strategy in our design.

However, simply assigning positive and non-negative region to a single scale will resulting ``unnecessary'' negative key-points. By unnecessarily negative we mean that in other scales, some key-points near to the center of an object may also contain some context information to cover the whole object. This can be well explained by the enlarged receptive field empowered by the downstream pathway and the CEM. One direct result of the unnecessarily negative key-points is the drop in recall because all these key-points are labeled as negative during training.

To solve this issue, we define a cosine Ground-truth Projection to maximize recall and speed up the training process. Figure \ref{fig:projection} illustrates the projection of the non-negative region across different feature maps when $conv6\_2$ is the best feature scale for the ground truth. The projection sizes of the non-negative region in neighboring feature maps are penalized with a cosine function, based on how far away they are from the best feature map $i, i \in [0, k-1] $. Suppose we have $ k $ feature maps, let $ d = |i-l|$  be the distance between current feature map $l$ and the best feature map $i$. We can calculate the penalizing factor $\varphi^l $ on feature map $l$ as follows:
\begin{equation}
\varphi^l = \max(\cos(\dfrac{\lambda d \pi}{2k}),0),
\end{equation}
where $\lambda$ decides to what extent the projection is penalized. We use $\lambda = 2.5$  in this paper. Our new  non-negative region $b^l_n$  can now be defined as:
\begin{equation}
b^l_n = (c_x,c_y,\varphi^l w^l_n,\varphi^l h^l_n).
\end{equation}
By using this cosine function we penalize  the non-negative region less on neighboring feature maps than on distanced feature map as shown in Figure \ref{fig:projection}. As a result, we are able to receive more responses from neighboring feature maps which further improves the recall performance of our model.

\subsection{Box Regression}
Unlike anchor-based methods which predict the offset by using reference anchor boxes, the output of our bounding box regression is the offset of the center point and size encoded by the stride of the feature map. In particular, given an object $b = (x_1 , y_1, x_2, y_2)$ with category $C_b$, the center position of the ground truth $(c^b_x, c^b_y)$ can be calculated as $ (\dfrac{x_2-x_1}{2}, \dfrac{y_2-y_1}{2}) $, with $ w^b = x_2 -x_1; h^b =y_2 -y_1$ being the width and height of the box. The offset vector $ v^l_{b,p_{ij}} =[\Delta c^{l}_x,\Delta c^{l}_y,\Delta w^l,\Delta h^l]$ for center point $  p_{ij} = (c^p_x,c^p_y) $ at  pixel location $(i,j)$ with stride $ s_l $ can be defined as:
\begin{equation}
\begin{aligned}
\Delta c^{l}_x &= \dfrac{(c^b_x-c^p_x)}{s_l}, & \Delta w^l &=\log(\dfrac{w^b}{s_l})\\
\Delta c^{l}_y &= \dfrac{(c^b_y-c^p_y)}{s_l}, & \Delta h^l &=\log(\dfrac{h^b}{s_l}).
\end{aligned}
\end{equation}
To remove overlapped bounding boxes, we apply IoU-based  Non Maximum Suppression (NMS) with a threshold of 0.1.

Similar to  Faster R-CNN \cite{DBLP:Faster-RCNN}, our localization loss for bounding box regression is Smooth L1 loss \cite{DBLP:Fast_rcnn}:
\begin{equation}
\mathcal{L}_{loc} = \frac{\beta}{N} \sum^{N}_k \text{SmoothL1}(v_k,\widehat{v}_k),
\end{equation} 
where $v_k,\widehat{v}_k$ and $N$ denote the predicted bounding boxes, ground truth bounding boxes and the number of positive labels respectively. In this paper, we have $ \beta =0.45 $.  


\subsection{Classification Loss}
\label{classification_loss}
Different from the situation we have in natural image modalities where objects are often dense (in terms of averaged number of objects in one image) and of regular sizes, in colonoscopy, polyps are often very sparse and small. This fact determines that we are facing an extremely small positive/negative sample ratio. Thus, we feel that Focal Loss fits better in our case compared with the Online Hard Negative Mining (OHEM) mechanism. 
However, we only apply Focal Loss \cite{lin2017focal} to negative samples while adopting cross entropy with a penalty term for positive samples. 
We assume center points closer to the ground truth centroid have a more precise view of an object and thus contribute more to the loss. 

In particular, the penalty weight of a certain positive center point is determined by its Euclidean distance to the ground truth centroid and the size of the object. We use an unnormalized 2-D Gaussian to generate the penalty weight. Formally, given an object $b = (c^b_x, c^b_y, w^b, h^b)$ and a point $ p_{ij} = (c^p_x,c^p_y) $ at pixel location $ (i,j)  $ within positive region, we can generate its weight $ \psi_{p_{ij}} $ by:
\begin{equation}
\psi_{p_{ij}}=\exp\big( -\dfrac{(c^b_x-c^p_x)^2 +( c^b_y-c^p_y)^2}{2\alpha(\max(w^b,h^b))^2}\big).
\end{equation}
We use $ \alpha = 1 $ in this paper. In sum, we will have the following loss defined for the classification end:
\begin{equation}
    \mathcal{L}_F(p_t) = -\alpha_t(1-p_t)^\gamma log(p_t)
\label{vfocal}
\end{equation}
\begin{equation}
\begin{aligned}
    \mathcal{L}_{cls} = 1/N_{pos}\sum_j^{N_{neg}}\mathcal{L}_F(p_{t_j}) 
    + 1/N_{pos}\sum_i^{N_{pos}}\psi_{i}\mathcal{L}_{CE}(p_{i}),
\end{aligned}
\label{tfocal}
\end{equation}
where $p_t = (p)^y(1-p)^{(1-y)}$, $y \in \{0,1\}$ is the ground truth. Finally we combined regression loss and classification loss to formulate our multitask loss $ \mathcal{L} =  \mathcal{L}_{loc}+ \mathcal{L}_{cls}$.

\section{Experiments}
We conduct our experiments mainly on GIANA \cite{MICCAI_2015}, CVC-CLINIC \cite{cvc_clinic} and ETIS-LARIB\cite{ETIS-LARIB} dataset. We use GIANA as the training set, CVC-CLINIC and ETIS-LARIB as the testing set.

\subsection{Data Preparation and Augmentation}
\textbf{GIANA}\cite{MICCAI_2015}, a database from MICCAI2017 endoscopic sub-challenge. The dataset contains four tasks, including Polyp detection, Polyp segmentation, Small Bowel Lesion Detection and Small Bowel Lesion Localization. For the polyp detection task, it contains 18 short videos for training and 20 short videos for testing. For the polyp segmentation task contains, it provides 300 images for training and 612 for testing and additionally more than 150 high definition images. However, the 612 testing image for the polyp segmentation task is identical to CVC-CLINIC.

To build our training set, we combined the datasets designed for Polyp detection and Polyp segmentation tasks from GIANA. Image frames are extracted from the 18 short videos. Segmentation annotations are converted to bounding boxes. As for the data augmentation purpose, we apply random rotation, zoom crop/expand, horizontal/vertical flip and distortions by using Augmentor\cite{Augmentor}. In the end, we obtain a training set of 25K images.

\textbf{CVC-CLINIC} \cite{cvc_clinic} contains 612 still frames from 29 endoscopic videos. Each image comes with manually labeled pixel-level ground truth by Computer Vision Center (CVC), Barcelona, Span. All these 612 images are used as our testing set. 

\textbf{ETIS-LARIB}\cite{ETIS-LARIB} contains 192 high resolution images with the resolution of 1225 $\times$ 996. Annotations are also provided at the pixel level. We convert these pixel-level annotations to bounding boxes for our detection task. We also evaluate our model on this dataset. 

\begin{table*}
	\begin{center}
		\begin{tabular}{|c|cccccc|c|c|c|c|}
			\hline		
			Experiment & FPN & CEM & OHEM &Focal Loss &Cosine Projection &Gaussian Penalty&Precision & Recall & F1-score & F2-score\\ 
			\hline\hline
			1 & & & & \checkmark  &\checkmark &\checkmark & 97.27  &93.65 &95.43 & 94.35\\ 
			2 & \checkmark& & & \checkmark  &\checkmark &\checkmark&  97.31 &	95.51&	96.33&	95.84\\
			3 &\checkmark &\checkmark & \checkmark &  &\checkmark &\checkmark&95.20&92.11&93.63&92.71 \\
			4 &\checkmark &\checkmark & & \checkmark&  &\checkmark&   98.56 & 95.36 & 96.93 &95.98\\
			5&\checkmark & \checkmark& & \checkmark  &\checkmark & &98.42  &  96.28 & 97.34 & 96.70 \\
			6 &\checkmark & \checkmark& & \checkmark  &\checkmark &\checkmark&  \textbf{99.36}&	\textbf{96.44}&	\textbf{97.88}&	\textbf{97.01}\\
		\hline
		\end{tabular}
		
	\end{center}
	\caption{Ablation study of each technical components. All the models use VGG16 as the backbone.}{\label{Tab:ablation_study}}
\end{table*}
\begin{table}[t]
	\begin{center}
		\begin{tabular}{|c|c|c|c|c|}
			\hline		
			Backbone &Precision & Recall & F1-score & F2-score\\ 
			\hline\hline
            ResNet-50 & 99.04  & 95.51 &97.24 &96.20\\
            ResNet-101 &98.40  & 95.36 &96.86 & 95.95 \\
			VGG16 & \textbf{99.36}&	\textbf{96.44}&	\textbf{97.88}&	\textbf{97.01}\\
		\hline
		\end{tabular}
	\end{center}
	\caption{Effectiveness of different backbones. For all these experiments, all technical components are added: FPN, CEM, Cosine projection and Gaussian penalty. 
	}{\label{Tab:ablation_study_backbone}}
\end{table}

\subsection{Training}
Our backbone network is initialized by the pre-trained VGG16 on ImageNet. All additional convolution layers added on top of the backbone network are initialized by the Xavier method. We train our model on NVIDIA V100 with a batch size of 64. We use SGD with 0.9 momentum, 0.0005 weight decay and initial learning rate of 0.001. During the training process, we apply cosine annealing  \cite{DBLP:cosine_learning_rate} for learning rate decay with an interval of 20 epochs. Following SSD, we also apply the online data augmentation photometric distortions and random sampling during training. 

\subsection{Evaluation metrics}
We follow the protocol of MICCAI2015 \cite{MICCAI_2015} challenge by using the precision, recall, F1, F2 score as the major evaluation metrics. We define the following terms to calculate the performance:

\textit{True Positive (TP)}: When the centroid of the predicted bounding box falls in the polyp ground truth, it will count as a True Positive detection.

\textit{False Positive (FP)}: When the centroid of the predicted bounding box falls outside of polyp ground truth, it will be viewed as False Positive.

\textit{False Negative (FN)}: When a polyp ground truth is not detected by our detector, it will count as a False Negative.

Note that, if multiple predicted bounding boxes fall within one polyp ground truth, only one TP will be counted. In sum, precision ($ pre $), recall ($ rec $), F1 and F2 scores are formulated as:
\begin{equation}
\begin{aligned}
pre &= \dfrac{TP}{TP+FP} , & rec &= \dfrac{TP}{TP+FN} \\
F1 &= \dfrac{2\times pre\times rec}{pre + rec}, & F2 &= \dfrac{5 \times pre \times  rec}{4 \times pre +rec}
\end{aligned}
\end{equation}

\begin{figure}[t]
	\begin{center}
		\includegraphics[width=0.9\linewidth,height=2in]{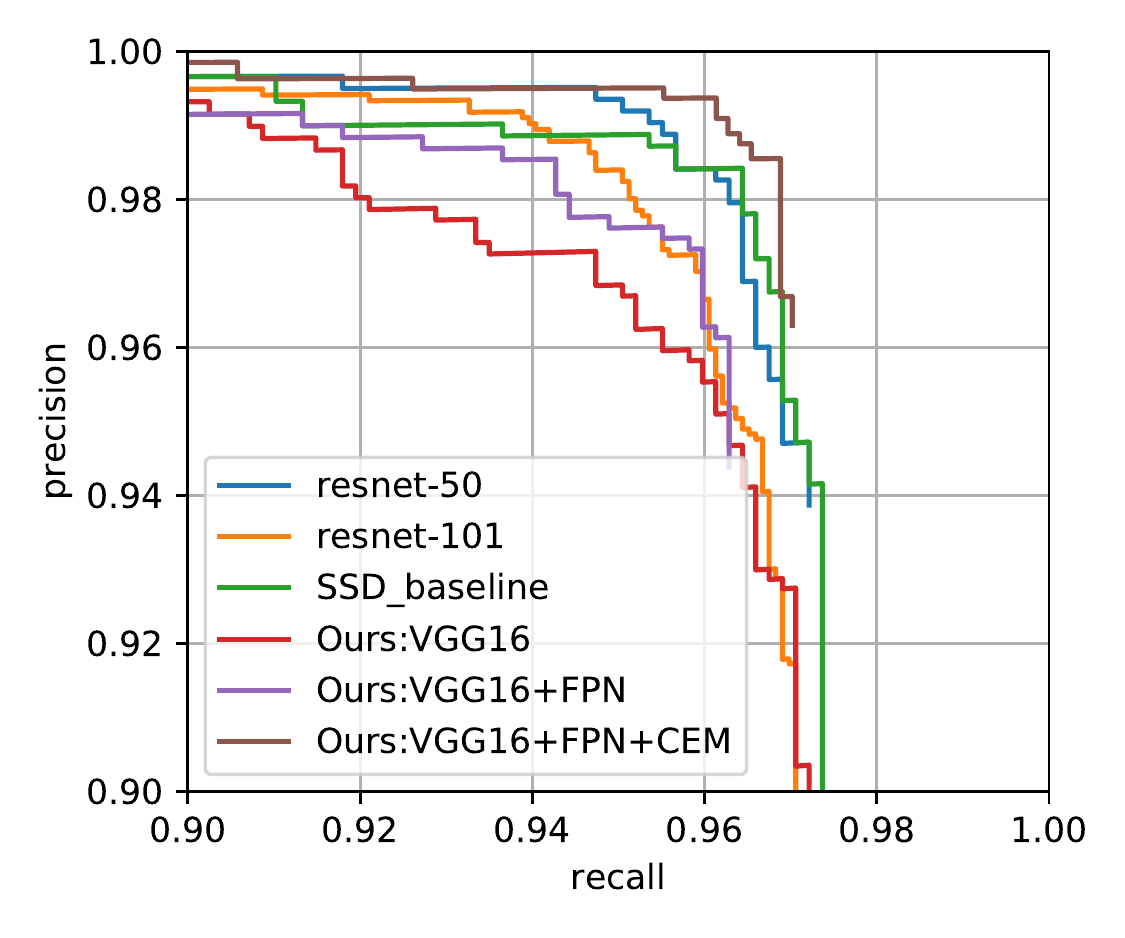}
	\end{center}
	\caption{ Precision Recall Curve for different networks.}
	\label{fig:result_graph}
\end{figure}

\subsection{Ablation Study}
We also conduct a full ablation study to isolate the effects of each technical component. All these models use VGG16 as the backbone and are evaluated with the CVC-Clinic dataset and results are summarized in Table \ref{Tab:ablation_study}.

\textbf{FPN improve Recall significantly}. In Experiment 1 (Table~\ref{Tab:ablation_study}), the downstream pathway is removed as well as the CEM. Therefore the network is similar to SSD. By comparing Experiment 1\&2, we find the recall rate is improved significantly. We attribute this to the enlarged effective receptive field from downstream features that carry more context information. This observation is further supported by the experiment on the Context Enhancement Module.

\textbf{Context Enhancement Module plays a critical role in improve the model}. By comparing Experiment 2\&6 in Table \ref{Tab:ablation_study}, the overall performance drops significantly, 2.1\% for recall and 0.9\% for precision, when CEM is removed. This is consistent with the effects of the Corner Pooling in CornerNet \cite{DBLP:cornernet} and in turn, confirms the importance of enriching context in anchor-free detectors. 

\textbf{Focal Loss works better than OHEM}. As we can observe from Table~\ref{Tab:ablation_study} that (Experiment 3\&6), Focal Loss brings in 4.1\%, 4.3\% increase in precision and recall over OHEM. Moreover, in practice, training would be much faster and more stable with Focal Loss than with OHEM.



\begin{table*}[t]
	\begin{center}
		\begin{tabular}{|l|c|c|c|c|c|c|c|c|c|}
			\hline
			Method & Testing Dataset & TP & FP & FN  & Precision & Recall & F1-score & F2-score & Inference time\\
			\hline\hline
			Y-Net\cite{DBLP:y-net}  &  ASU-MAYO* &3582 &513 &662 &87.4 &84.4 &85.9 & 85.0 & N/A\\
			RYCO\cite{rest-yolo} &  ASU-MAYO* &3087&398&1226&88.6&71.6&79.2&74.4&N/A\\
			\hline
			ASU \cite{Polyp_ASU}&  CVC-Clinic-test & N/A &N/A &N/A&97&85.2&90.8&87.4&N/A\\
			CVC-CLINIC\cite{cvc_clinic} &  CVC-Clinic-test & N/A &N/A &N/A&83.5&83.1&83.3&83.2&N/A\\
			\hline
			OUS\cite{MICCAI_2015} & ETIS-LARIB & 131 & 57 & 77 & 69.7 & 63.0 & 66.1 & 64.2 &N/A\\
			RCNN-Mask\cite{RCNN-Mask}&  ETIS-LARIB  &167 & 62 & 41 & 72.93  & 80.29  &76.43  & 78.70 &N/A \\
			FRCNNPL\cite{fatser-rcnn-with-post-learning}  &  ETIS-LARIB &167 &26 &41 &86.5 &80.3 &83.3 & 81.5 & 5.0 FPS\dag\\ 
			\textbf{AFP-Net (ours)}&ETIS-LARIB &168 & 21 & 40 & \textbf{88.89}  &  \textbf{80.77} & \textbf{84.63} &\textbf{ 82.27} & \textbf{ 52.6 FPS} \\
			\hline
			Mask-RCNN \cite{MASK_RCNN_POLYP_DETECTION} & CVC-Clinic-train & N/A & N/A &N/A&83.49&92.95&87.96&90.89 & N/A\\
			Faster-RCNN \cite{polyp-faster-rcnn} & CVC-Clinic-train & 523 & 81 &8&86.6&\textbf{98.5}&92.2&95.9 & N/A\\
			CenterNet-104 & CVC-Clinic-train & 603 &30 &43 &95.27 &93.35 &94.30 & 93.73 & 7.8 FPS\\ 
			SSD-baseline** &  CVC-Clinic-train  & 618 & 8 & 28 & 98.72  & 95.67 & 97.17 & 96.26 &37.0 FPS\\ 
			\textbf{AFP-Net (ours)} & CVC-Clinic-train & 623  & 4 &23&\textbf{99.36}&  96.44&\textbf{97.88}&\textbf{97.01}&\textbf{52.6 FPS}\\ 
			\hline
		\end{tabular}
	
	\end{center}
	\caption{Results of proposed method compare to others. * dataset no longer available. ** baseline SSD (FPN + CEM + SSD Anchors) with input size of 320. $\dag$ runs with a NVIDIA GTX TITAN X.}{\label{Tab:result_compare}}
\end{table*}

\begin{figure*}[t]
	\centering
	\begin{subfigure}[b]{0.225\linewidth} %
		\centering
		\includegraphics[width=1\textwidth]{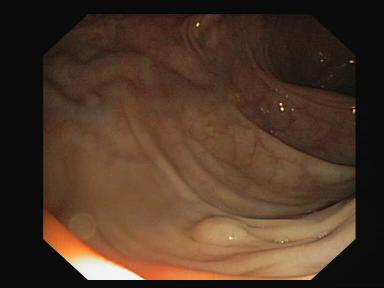}%
	\end{subfigure}%
    \begin{subfigure}[b]{0.225\linewidth} %
		\centering
		\includegraphics[width=1\textwidth]{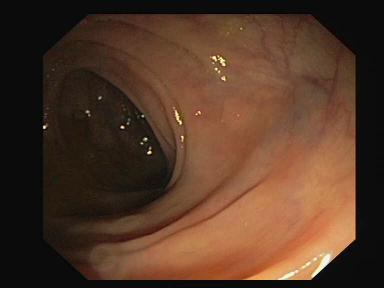}%
	\end{subfigure}%
	\begin{subfigure}[b]{0.225\linewidth} %
		\centering
		\includegraphics[width=1\textwidth]{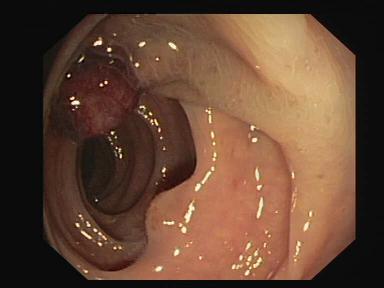}%
	\end{subfigure}%
	\begin{subfigure}[b]{0.225\linewidth} %
		\centering
		\includegraphics[width=1\textwidth]{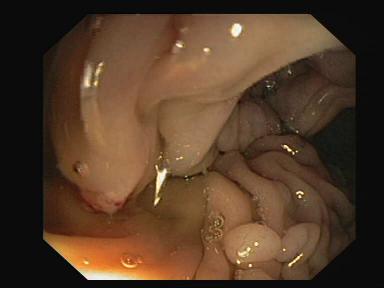}%
	\end{subfigure}%

    \begin{subfigure}[b]{0.225\linewidth}%
    \centering
    \includegraphics[width=1\textwidth]{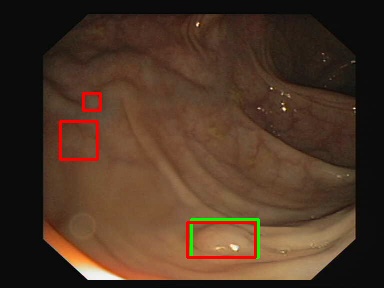}%
    \caption{}
    \label{fig:hard_case1}
    \end{subfigure}%
	\begin{subfigure}[b]{0.225\linewidth}%
	\centering
	\includegraphics[width=1\textwidth]{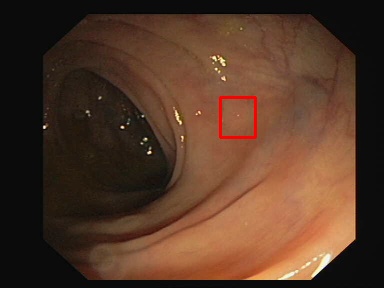}%
	    \caption{}
		\label{fig:hard_case3}
    \end{subfigure}%
    \begin{subfigure}[b]{0.225\linewidth}%
	\centering
	\includegraphics[width=1\textwidth]{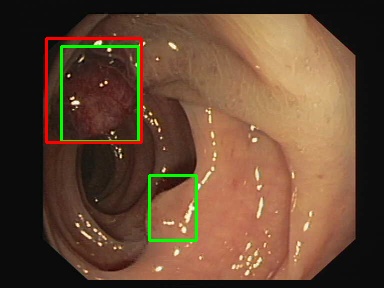}%
	    \caption{}
		\label{fig:false_positive1}
    \end{subfigure}%
    \begin{subfigure}[b]{0.225\linewidth}%
	\centering
	\includegraphics[width=1\textwidth]{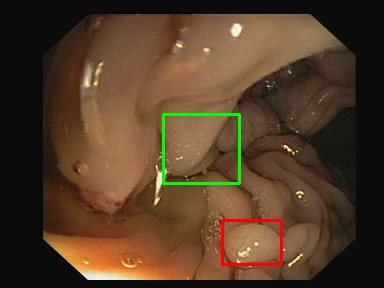}%
	    \caption{}
		\label{fig:false_positive2}
    \end{subfigure}%

	\caption{Examples of False Negative (a \& b) and False Positive (c \& d). The first row shows the raw input images while the second row presents the corresponding detection results. The green rectangles are our predicted bounding boxes. The red rectangles are the ground truth boxes.}
	\label{fig:failed_hard_case} 
\end{figure*}

\textbf{Cosine projection results in performance boost}. In Experiment 4, we remove the cosine projection, then a ground truth would only be assigned with one feature scale. Comparing Experiment 4\&6 in Table \ref{Tab:ablation_study}, we can observe that the recall rate is increased by a considerable margin.


\textbf{Gaussian penalty improves precision}. In order to investigate the effectiveness of the Gaussian penalty, we train one network without it (Experiment 5). As we can see by comparing Experiment 5\&6, the Gaussian penalty results in 0.94\%, 0.18\% increase in precision and recall respectively. 

We also explored the effect of different backbones in our model (Table \ref{Tab:ablation_study_backbone}), finding that VGG seems to work better than ResNets. In the meantime, to give a better view of the performance, we show the ROC curve with respect to precision and recall in Fig. \ref{fig:result_graph}.

\subsection{Overall Performance}
We compared our proposed method with other models reported in the MICCAI2015 challenge on polyp detection \cite{MICCAI_2015} and recent work \cite{polyp-faster-rcnn,DBLP:y-net,fatser-rcnn-with-post-learning}. We run our test on CVC-Clinic training dataset and ETIS-LARIB dataset, with an input size of 320. We set the IoU threshold to 0.1 for the non-maximum suppression. All results are summarized in Table \ref{Tab:result_compare}.

We refer to our model as AFP-Net (Experiment 6 in Table \ref{Tab:ablation_study}). Our proposed method outperforms all previous approaches in terms of F1 and F2 scores on both testing datasets. In the meantime, we also compared our approach with another anchor-free design (CenterNet), showing the superior performance of our design. Note that our anchor-free detector even outperforms its anchor-based baseline model SSD-baseline (FPN + CEM + SSD anchors) where only the anchor part is different. 

We test all models on a NVIDIA RTX-2080TI with an input size of 320 to investigate the inference speed (FPS, frame per second) of our models. Among all anchor-based models, the ``SSD-baseline" (differ with AFP-Net only in the anchor design) represents the fairest and direct reference. As shown in Table \ref{Tab:result_compare}, our model gives a speed boost around 30\%. This speed boost is consistent when comparing with other anchor-based designs such as FRCNNPL \cite{fatser-rcnn-with-post-learning}. However, we must note that in FRCNNPL, the inference time is evaluated with a NVIDIA GTX TITAN X which is slightly slower than our GPU. Nevertheless, we are still confident to claim that our model runs faster, given the large gap in inference time (52.6 FPS vs. 5 FPS). On the other hand, to compare with other anchor-free designs, we retrained a CenterNet. As shown in Table \ref{Tab:result_compare}, the speed advantage of our model still holds. 


\subsection{Visualization} 
We visualize some of the hard cases in Fig. \ref{fig:failed_hard_case} where our model can make a mistake. Our predicted boxes are marked as green rectangles while the ground truth for each image is marked as red rectangles. As we can see the missed polyps in Figure \ref{fig:hard_case1} and \ref{fig:hard_case3} are very challenging cases because they share similar texture and shapes with normal mucosae. This is partially due to the view angle during image capturing. When the light hits on top of a polyp directly, the polyp may blend into the background and becomes difficult to detect. Figure \ref{fig:false_positive1} and  \ref{fig:false_positive2} show some samples of false positive detection. Normal mucosae with some fold structures are very hard to be distinguished from polyps, especially the early stage of polyps, as they can share the same texture as shown in Figure \ref{fig:false_positive2}.

\section{Conclusion and Future work}
In this paper, we proposed a novel anchor-free polyp detector. It is faster than the anchor-based design while achieving state-of-the-art performance. In addition to the improved performance, we also remove the hassle of manually fine-tuning anchor related hyper-parameters. 

Enriching context information is critical for anchor-free detectors. Feature Pyramid, Context Enhancement Module all contribute in this way. At the Loss end, Focal Loss, distance based Gaussian Penalty and our proposed cosine ground truth all play important roles in improving the performance. With all these technical components, the potential recall drop caused by removing anchors are well compensated and we achieved the state-of-the-art performance.


We believe our cosine ground truth projection and Gaussian penalty will provide a vital building block for future anchor-free design. In our future work, we intend to further improve the classification strategies for small flat polyps that are very hard to be distinguished from normal mucosae.

\bibliographystyle{ieee}
\bibliography{egbib}

\end{document}